%Paper: astro-ph/9407067
%From: TUROLLA@padova.infn.it
%Date: Fri, 22 Jul 1994 14:57:13 +0200 (WET-DST)

\magnification=1200
\hyphenpenalty=2000
\tolerance=10000
\hsize 14.5truecm
\hoffset 1.truecm
\openup 5pt
\baselineskip=24truept
\font\titl=cmbx12
%     \der{f}{x}    ====>  derivata parziale df/dx
\def\der#1#2{{\partial#1\over\partial#2}}
%     \ders{f}{x}{y}  ====>  derivata parziale seconda d2f/dxdy

%      \derss{f}{x}   ====>  derivata parziale seconda d2f/dx2
\def\derss#1#2{{\partial^2#1\over\partial#2^2}}

\def\Tg{T_\gamma}

\def\kes{\kappa_{es}}
\def\kff{\kappa_{ff}}
\def\ref{\par\noindent\hangindent 20pt}
\def\refig{\par\noindent\hangindent 15pt}
\def\mincir{\raise -2.truept\hbox{\rlap{\hbox{$\sim$}}\raise5.truept
\hbox{$<$}\ }}
\def\magcir{\raise -4.truept\hbox{\rlap{\hbox{$\sim$}}\raise5.truept
\hbox{$>$}\ }}
\def\rho{\varrho}

\null
\vskip 1.truecm
\centerline{\titl X--RAY SPECTRA FROM NEUTRON STARS}
\smallskip
\centerline{\titl ACCRETING AT LOW RATES}
\vskip 1.truecm
\centerline{Luca Zampieri $^1$, Roberto Turolla $^2$, Silvia Zane $^1$
and Aldo Treves $^1$}
\bigskip
\centerline{$^1$ International School for Advanced Studies, Trieste}
\centerline{Via Beirut 2--4, 34014 Trieste, Italy}
\medskip
\centerline{$^2$ Department of Physics, University of Padova}
\centerline{Via Marzolo 8, 35131 Padova, Italy}
\vfill
\bigskip\bigskip
\beginsection ABSTRACT

The spectral properties of X--ray radiation produced in a static
atmosphere around a neutron star accreting at very low rates are
investigated. Previous results by Alme \& Wilson (1973) are extended
to the range $10^{-7}\leq L/L_{Edd}\leq 10^{-3}$ to include the typical
luminosities, $L\sim 10^{31}-10^{32} \ {\rm ergs\, s^{-1}}$, expected from
isolated neutron stars accreting the interstellar medium. The emergent
spectra show an overall hardening with respect to the blackbody at the
neutron star effective temperature in addition to a significant excess over
the Wien tail. The relevance of present results in connection with the
observability of low--luminosity X--ray sources is briefly discussed.
\bigskip\bigskip
\noindent
{\it Subject headings:\/} accretion, accretion disks \ -- \  radiative
transfer \ -- \  stars: neutron \ -- \ X--rays: stars\hfill\break
\vfill\eject

\beginsection 1. INTRODUCTION

Since the early seventies large theoretical efforts have been devoted to
investigate the properties of radiation produced by accretion onto neutron
stars, in the attempt to model the observed spectra of galactic X--ray sources.
Up to now, the interest mainly focussed on emergent spectra for luminosities in
the range $\sim 10^{35}-10^{38} \ {\rm ergs\, s^{-1}}$, since most of observed
X--ray binaries have $L\magcir 10^{34} \ {\rm ergs\, s^{-1}}$. Much fainter
sources may, nevertheless, exist. It was suggested long
ago (Ostriker, Rees, \& Silk 1970), in fact, that
old isolated neutron stars (ONSs) no longer active as pulsars,
may show up as very weak, soft X--ray sources ($L \sim 10^{31}-10^{32}\,
{\rm ergs\, s^{-1}}$) if they accrete the interstellar medium. The capabilities
of present instrumentation on board X--ray satellites like ROSAT, make now
possible the observation of such low--luminosity sources (e.g. Treves \&
Colpi 1991, Blaes \& Madau 1993).

ONSs may have low magnetic fields in which case they would accrete spherically.
%and high velocity relative to the ISM.
Spherical accretion
onto an unmagnetized neutron star was firstly considered by Zel'dovich \&
Shakura (1969, ZS in the following) and, in more detail, by Alme \& Wilson
(1973, AW), who concentrated, however, on high--luminosity solutions.
The lack of results for $L\mincir 10^{-3} L_{Edd}$ led us to
reconsider the problem. We compute
the spectrum emerging from a static spherical atmosphere
surrounding an accreting neutron star in the case of very low rates,
extending previous calculations by AW to $L\sim 10^{-7}\, L_{Edd}$.
We found that the resulting spectral distributions are significantly harder
than a blackbody at the star effective temperature. Our results may lead to
reconsider the present estimates of the observability of ONSs,
which were based on the assumption of a Planckian spectrum.

\beginsection 2. THE MODEL

We consider a non--rotating, unmagnetized neutron star which undergoes
spherical accretion and is surrounded by a spherical, static atmosphere;
the envelope material is assumed to be pure hydrogen. As the
accretion flow penetrates into the atmosphere, protons are decelerated by
Coulomb collisions and/or plasma interactions, their bulk kinetic energy is
transferred to electrons, and is finally converted into electromagnetic
radiation via free--free emission. The input physics of our model essentially
coincides with that used in previous studies on this subject. Equilibrium
solutions were firstly provided by ZS for the frequency--integrated case, and
by AW for the complete transfer problem.
These investigations concentrated on the study of high--luminosity regimes
($L \magcir 10^{36} \, {\rm ergs\, s^{-1}}$)
and revealed the fundamental role played by
Comptonization of primary bremsstrahlung photons in establishing the thermal
balance of the emitting gas and the properties of the emerging radiation.
The possible existence of a new ``hot'' solution, in which Comptonization
is dominant, was recently discussed by Turolla, Zampieri, Colpi \& Treves
(1994, Paper I). A detailed modelling of the interactions between the
impinging flow and the static envelope is exceedingly complicated (see e.g.
Bildsten, Shapiro \& Wasserman 1992), also because a collisionless,
standing shock can form,
as originally suggested by Shapiro \& Salpeter (1973). Following both ZS and
AW (see also Paper I), we circumvent this problem assuming that the proton
stopping length is a free parameter of our model, together with the total
luminosity $L_\infty$, measured by an observer at infinity. Denoting by
$\rho$ the gas density,  the column density of the atmospheric material is
$y = \int_R^\infty\rho\, dR$; the value $y_0$ corresponding to the proton
stopping length
was estimated by ZS to be in the range $5\mincir y_0\mincir 30
\, {\rm g \, cm^{-2}}$ when
Coulomb collisions are dominant. As it was shown in Paper I, no significant
expansion occurs in both ``cold'' and ``hot'' envelopes, so we
treat the radial coordinate as a constant, equal to the neutron star radius
$R_*$. The heat injected per unit time and mass in the envelope
is calculated using the expression for the energy loss rate due to Coulomb
collisions of super--thermal protons (see Bildsten {\it et al.\/}),

$$W_h = \cases{\displaystyle{{L_\infty}\over{8\pi R_*^2y_0y_G}}
{{1 + v_{th}^2/v_i^2}\over{[1 - (1-v_{th}^4/v_i^4)(y/y_0)]^{1/2}}}&
$y\leq y_0$\cr &\cr
0 & $y > y_0$\cr}\eqno(1)$$
where $v_i^2 = c^2 (1 - L_\infty/L_{Edd})/3$ is the ``modified'' free--fall
velocity, $v_{th}^2 = 3 k T/m_p$ is the proton thermal velocity,
$y_G = (1 - 2GM_*/c^2R_*)^{1/2}$ is the gravitational redshift factor
in the Schwarzschild spacetime and $M_*$ is the star mass. We note
that, owing to the gravitational redshift, the total luminosity seen by a
distant observer, $L_\infty$, is related to the local luminosity at the top
of the atmosphere by $L_\infty = y_G^2 L(0)$.

The run of pressure $P$, temperature $T$, monochromatic radiation energy
density $U_\nu$ and flux $F_\nu$ (both measured by the local observer) are
obtained solving the
hydrostatic and the energy balance for a completely ionized, perfect hydrogen
gas (ZS, Paper I) coupled to the first two frequency--dependent transfer
moment equations in the Eddington approximation (AW; Nobili, Turolla \&
Zampieri 1993, hereafter NTZ)

$$\eqalignno{
 & {{dP}\over {dy}} = {{GM_*}\over{y_G^2 R_*^2}}\left(1 -
{{\kappa_1}\over{\kappa_{es}}}{{y_G L}
\over { L_{Edd}}}\right)  & (2)\cr
 & {W\over c} = {\kappa}_P\left( aT^4 - {{\kappa_0}\over{\kappa_P}} U\right) +
4\kes U{{KT}\over{m_ec^2}}\left(1 - {{\Tg}\over T}\right) - {{c\kappa_{es}}
\over{8\pi m_e}}\int\left({{U_\nu}\over\nu}\right)^2 d\nu\, & (3)\cr
 & {{F_\nu}\over{c U_\nu}} \left[ \der{\ln F_\nu}{y}
- {{GM_*}\over{c^2\rho y_G^2R_*^2}} \left(1 - \der{\ln F_\nu}{\ln \nu}
\right)
- {{2}\over{\rho R_*}} \right] = - {{s_\nu^0}\over{y_G U_\nu \rho}}
 & (4)\cr
 & {1\over 3}\der{\ln U_\nu}{y} - {{GM_*}\over{c^2\rho y_G^2R_*^2}}
 \left(1 - {1\over 3}\der{\ln U_\nu}{\ln \nu}\right)
= - {{s_\nu^1}\over{y_G U_\nu \rho}}.&(5)\cr}$$
Here $U=\int U_\nu\, d\nu$, $L=4\pi R_*^2\int F_\nu\, d\nu$,
$L_{Edd}=4\pi GM_*c/\kes$; $\kes$, $\kappa_P$, $\kappa_0$ and
$\kappa_1$ are the scattering, Planck, absorption
and flux mean opacities, and the radiation temperature $\Tg$ is defined by
$$\Tg = {h\over{4k}}{{\int\nu U_\nu\, d\nu}\over{\int U_\nu\, d\nu}}\, .
\eqno(6)$$
$W = W_h - W_c$ represents that part of the injected heat, $W_h$, which is
effectively converted into electromagnetic radiation within the atmosphere.
$W_c$ mimics the possible presence of other forms of energy transport
(like convection and electron conduction) which are not treated
in detail here (see AW for comparison); the actual form of $W_c$ is discussed
later on.
The first two moments of the source function, $s_\nu^0$ and $s_\nu^1$,
account for the exchange of energy and momentum
between electrons and photons and, for the radiative processes we are
considering, bremsstrahlung and electron scattering, they have the form
(NTZ)
$$\eqalignno {&{{s_\nu^0}\over{U_\nu }} =
\kes\rho\left\{{{kT}\over{m_ec^2}}
\left[\derss{\ln U_\nu}{\ln \nu} + \left(\der{\ln U_\nu}{\ln \nu}
 + {{h\nu}\over{kT}}
-3\right)\der{\ln U_\nu}{\ln \nu} +  \right.\right. \cr
&\left.\left. + {{h\nu}\over{kT}} +
{{c^3 U_\nu}\over{4\pi kT \nu^2}}\left(\der{\ln U_\nu}{\ln \nu}
-1\right)\right]
+ {\kff\over\kes}\left({{4 \pi B_\nu(T)}\over{U_\nu c}}-1\right)\right\}&(7)\cr
}$$

$$ s_\nu^1 = -\left(\kes + \kff\right)\rho {{F_\nu}\over {c }}\, , \eqno(8)$$
where $B_\nu(T)$ is the Planck function. The free--free opacity for
a completely ionized hydrogen gas is
$$\kff = 1.318 \times 10^{56} \rho T^{-1/2} {{1 - e^{-h\nu / kT}}\over
\nu^3} \, {\bar g(\nu, T)} \, {\rm cm^2 \, g^{-1}} \, .$$
and a functional fit to Karzas \&
Latter's (1961) tables was used for the velocity--averaged Gaunt factor.
Since equations (4) and (5) define a second order elliptic operator,
conditions must be prescribed on the entire
boundary of the integration domain and their form is discussed in
NTZ. In particular, we assume that diffusion holds in the deeper layers
where LTE is certainly attained and this automatically fixes
the luminosity at the inner boundary, $L_{in}$. If all the energy is
supplied by accretion, the total radiative flux that crosses the inner
boundary must equal the heat transported inward by non--radiative processes,
that is to say

$$W_c = {L_{in}\over{8\pi R_*^2y_0y_G}}
{{1 + v_{th}^2/v_i^2}\over{[1 - (1-v_{th}^4/v_i^4)(y/y_0)]^{1/2}}}
\quad \quad \quad \quad 0 \leq y \leq y_{in} \, .$$

%$W_c = y_GL_{in}/4\pi R_*^2y_0$.

Although the total radiation energy density and luminosity are just the
integrals of $U_\nu$ and $4\pi R_*^2 F_\nu$ over frequency, we found
numerically more convenient to derive them from the first two gray moment
equations
$$\eqalignno{
 &{{dL}\over{dy}} = - {{4\pi R_*^2 W}\over y_G} \, & (9)\cr
 &{1\over 3}{{dU}\over{dy}} =  \kappa_1 {L\over{4\pi R_*^2 c y_G}}\, .
 & (10)\cr}$$
Equation (9) gives trivially
$$L = \cases{\displaystyle {{L_\infty}\over{y_G^2}}- \left(
{{L_\infty}\over{y_G^2}} - L_{in}\right)
{{1 - [1 - (1-v_{th}^4/v_i^4)(y/y_0)]^{1/2}}\over{1 - v_{th}^2/v_i^2}}
&  $y\leq y_0$\cr
&\cr
L_{in} & $y > y_0$\cr}\eqno(11)$$
with the condition $L(y_{in}) = L_{in}$, while equation (10) is integrated
numerically
along with the system (2)--(5), imposing $U(0) = 2F(0)/c\, $; pressure vanishes
at the top of the atmosphere, $P(0) = 0$.

\beginsection 3. RESULTS

Equations (2)--(5) and (10) were solved numerically by means of a finite
differences
relaxation scheme (NTZ) on a logarithmic grid of 50 frequency bins $\times$
100 depth zones. The adimensional frequency $x = h\nu/kT_*$ was used,
$T_* =T(y_{in})$, and the integration range was typically $-0.7< \log x < 1.1$,
$-7.6< \log y < \log y_{in}$, with $y_{in}$ marginally smaller than $y_0$.
A typical run required $\sim$ 15 minutes of CPU time on an IBM RISK/6000.
Two sets of models were computed, both with $R_* = 12.4$ km, $M_* = 1.4
M_\odot$:  $y_0 = 20 \ {\rm g\, cm^{-2}}$ and luminosities in the range
$10^{-7}\leq L_\infty/L_{Edd}\leq 0.2$,  $y_0 = 5 \ {\rm g\, cm^{-2}}$ and
$10^{-7}\leq L_\infty/L_{Edd}\leq 10^{-5}$. Our numerical
method should guarantee a fractional accuracy better
than $1\%$ on all the variables. As a further check,
the total luminosity, given by equation (11), was compared with the
numerical integral of $F_\nu$ over the frequency mesh at each depth:
agreement was always
better than $10\%$. We have also verified that our solutions with
$L_\infty\magcir 10^{-2}L_{Edd}$ reproduce almost exactly those computed by AW.
For low--luminosity models, which we are mainly interested in, particular care
must be used to handle properly the absorption and flux mean opacities, since
the envelope thermal balance depends entirely on the free--free integrated
source term and the radiation spectrum becomes very nearly Planckian in the
deeper layers.

Results are summarized in figures 1, 2 and 3, where the emergent spectra and
the temperature profiles are plotted for different values of $L_\infty$. In all
these models it is $y_0 = 20 \ {\rm g\, cm^{-2}}$; solutions with $y_0 = 5 \
{\rm g\, cm^{-2}}$ show the same qualitative behaviour. A quite unexpected
feature emerging from figures 1 and 2 is that the spectral shape deviates more
and more from a blackbody as $L_\infty$ decreases. The model with $L_\infty =
2.25\times 10^{-2} L_{Edd}$ is, in fact, quite Planckian in shape (see also
AW), showing only a moderate hard excess. On the contrary, solutions with
$L_\infty < 10^{-4} L_{Edd}$ are characterized by a very broad maximum and by a
slow decay at high energies. Comptonization is relatively important for $L
\magcir 10^{-2} L_{Edd}$, similarly to what happens in X--ray burster
atmospheres (see e.g. London, Taam, \& Howard 1986). For less
luminous models, however, non--conservative scatterings play essentially no
%both in the thermal balance and
role in the formation of the spectrum, as it should be expected since the
temperature, and hence the Compton parameter, becomes lower. As can be seen
from figure 3, the temperature profile is nearly adiabatic in the inner layers
where the gas is optically thick to true emission--absorption at all
frequencies; for $L \mincir 10^{-5} L_{Edd}$ an intermediate, isothermal region
is present. The sudden increase of $T$ in the external layers is due to the
heating produced by the incoming protons, balanced
mainly by Compton cooling at low densities. The temperature ``shock'' moves
at very low values
of the column density and is nearly out of our depth mesh for $L_\infty =
10^{-7} L_{Edd}$.

Although the hard excess present in our low--luminosity spectra may be of
interest as far as predictions on the observability of ONSs are concerned,
more important, in this respect, seems to be the comparison of the actual
emerging spectrum with the blackbody at the neutron star effective temperature,
$B_\nu(T_{eff})$, $T_{eff} = [L_\infty/(4\pi R_*^2\sigma )]^{1/4}$, which was
assumed to be the emitted spectrum in all previous investigations
(e.g. Treves \& Colpi 1991, Blaes \& Madau 1993).
%In figure 3 the spectra of the two models with $y_0 = 20 \ {\rm g\, cm^{-2}}$
%and $L=10^{-6}\, ,10^{-7}L_{Edd}$ are shown together with the correspondent
%Planck functions at $T_{eff}$.
It is apparent from figure 2 that model spectra with $L \mincir 10^{-5}
L_{Edd}$ are substantially harder than the blackbody at the star effective
temperature.
The spectral hardening can be quantified introducing a hardening ratio
$$\gamma = {{\Tg}\over {\Tg [B_\nu(T_{eff})]}} \, , \eqno(12)$$
where the radiation temperature $\Tg$ is defined in equation (6) and
$\Tg [B_\nu(T_{eff})] = 0.96 T_{eff}$. This differs from the usual definition,
$\gamma = T_{col}/T_{eff}$, where $T_{col}$ is the color temperature,
because our spectra are not always well fitted by a blackbody.
For $y_0 = 20 \ {\rm g\, cm^{-2}}$, $\gamma$ steadily increases from
$\sim 1.5$ (value typical of X--ray bursters in the static phase),
for $L\sim 10^{-2}-10^{-3}L_{Edd}$,
up to $\sim 2.5$ for $L\sim 10^{-6}-10^{-7}L_{Edd}$ (see table 1).

\beginsection 4. DISCUSSION AND CONCLUSIONS

The significant deviation of low--luminosity spectra from
a Planckian equilibrium distribution could appear unexpected,
%rather surprising at a first glance,
since radiation is in LTE in a medium where the
scattering depth is always much less than the absorption one.
The source function should be Planckian
and the emergent spectrum, formed at the thermal photosphere, should coincide
with $B_\nu(T_{eff})$. However, if the atmosphere
develops smooth temperature and density gradients in layers where
the medium becomes optically thin to free--free, the differential nature
of absorption opacity plays an important role. High--frequency photons
decouple in the deeper, hotter layers and then propagate freely
to infinity, contributing to the high--energy part of the
emergent spectral flux. At large enough frequencies,
the observed shape of the spectrum turns out
to be a superposition of planckians at different temperatures. This result
resembles closely that of standard accretion disks, where the
emergent spectrum shows a broad plateau due to the combined, thermal emission
of rings at different temperatures.

Our present result that
low--luminosity spectra are harder than a blackbody is consistent with
the previous finding by Romani (1987), who computed model atmospheres for
cooling neutron stars. Although he considered a quite different physical
scenario, an atmosphere in radiative energy equilibrium illuminated from below,
the free--free opacity in his cool, He models ($T_{eff}\sim 3\times 10^5$ K)
acts much in the same way as in our faint solutions, producing a
hardening of the spectrum. Both in Romani's and in our analysis the effects of
the neutron star magnetic field were ignored. An insight on the role of a
strong {\bf B} field, $\sim 10^{12}$ G, on radiative transfer was recently
provided by Miller (1992), who showed that departures from a blackbody
become less pronounced, since opacity is increased by magnetic effects.
Finally we note that the assumption of a pure hydrogen chemical composition
used here, is not entirely ad hoc. In fact, contrary to what happens in
equilibrium atmospheres, such as those considered by Romani and Miller who
allowed for different compositions, it is
likely that metals are destroyed in the accretion flow (Bildsten {\it et
al.\/}), leaving just a hydrogen envelope.

As we already stressed, a motivation for studying the spectral
properties of X--ray radiation coming from neutron stars accreting at low
rates, stems from the possible detection of isolated
objects fed by the interstellar gas. Their expected luminosities, $\sim 10^{31}
\ {\rm ergs\, s^{-1}}$, could be within reach of satellites like Einstein and
ROSAT (see Treves \& Colpi; Blaes \& Madau; Colpi, Campana \& Treves 1993;
Madau \& Blaes 1994).
Very recently Stocke {\it et al.\/} (1994) pointed out
that one of the objects in the Einstein Extended Medium Sensitivity Survey
may be actually an ONS, consistently with the original suggestion by Treves \&
Colpi. The knowledge of the emitted spectrum is fundamental
in estimating the observability of any X--ray source and
a detailed analysis of the consequences of our results on the detectability
of ONSs with ROSAT will be presented in forthcoming paper (Colpi {\it et al.\/}
1994). Here we point out that synthetic spectra, being significantly
harder than the blackbody at $T_{eff}$, may indeed increase the chances of
detection. In table 1 we have listed the  ratios of the computed flux above
0.1 keV to the blackbody one for various luminosities; the threshold of 0.1
keV was suggested by the sensitivity of ROSAT. The solutions with $L = 10^{-7}$
and $10^{-5}\, L_{Edd}$ can be taken as representative of the typical
luminosities expected from ONSs embedded in the average ISM or in Giant
Molecular Clouds (see Colpi, Campana \& Treves).
As can be seen from the table, the ratio becomes
larger than unity and the flux above 0.1 keV is from $\sim 10\%$ to $\sim 40\%$
larger than the blackbody one for $10^{32}\magcir L\magcir 10^{31} \ {\rm
ergs\, s^{-1}}$.

Our present models could be relevant also to Soft X--ray Transients in
quiescence, such as Aql X--1 (Verbunt {\it et al.\/} 1993), or in connection
with low--luminosity
globular cluster X--ray sources which emit a luminosity $\sim 10^{-4}\,L_{Edd}$
(see e.g. Hertz, Grindlay, \& Bailyn 1993 and references therein). These still
mysterious objects could be either accreting white dwarfs (e.g. cataclysmic
variables) or neutron stars in binary systems. Their spectrum, which is still
poorly known, could be compared with the results of our model, and deviations
from a blackbody could be an important clue in discriminating their physical
nature.

\beginsection ACKNOWLEDGEMENTS

We thank an anonymous referee for some helpful comments.

\beginsection REFERENCES

\ref{Alme, M.L., \& Wilson, J.R. 1973, ApJ, 186, 1015 (AW)}
\ref{Bildsten, L., Salpeter, E.E., \& Wasserman, I. 1992, ApJ, 384, 143}
\ref{Blaes, O., \& Madau, P. 1993, ApJ, 403, 690}
\ref{Colpi, M., Campana, S., \& Treves, A. 1993, A\&A, 278, 161}
\ref{Colpi, M., Treves, A., Turolla, R., Zampieri, L., \& Zane, S. 1994,
in preparation}
\ref{Hertz, P., Grindlay, J.E., \& Bailyn, C.D. 1993, ApJ, 410, L87}
\ref{Karzas, W.J., \& Latter, R. 1961, ApJS, 6, 167}
\ref{London, R.A., Taam, R.E., \& Howard, W.E. 1986, ApJ, 306, 170}
\ref{Madau, P., \& Blaes, O. 1994, ApJ, 423, 748}
\ref{Miller, M.C. 1992, MNRAS, 255, 129}
\ref{Nobili, L., Turolla, R., \& Zampieri, L. 1993, ApJ, 404, 686 (NTZ)}
\ref{Ostriker, J.P., Rees, M.J., \& Silk, J. 1970, Astrophys. Letters, 6, 179}
\ref{Romani, R.W. 1987, ApJ, 313, 718}
\ref{Shapiro, S.L., \& Salpeter, E.E. 1973, ApJ, 198, 761}
\ref{Stocke, J.T., Wang, Q.D., Perlman, E.S., Donahue, M., \& Schachter, J.
1994, preprint}
\ref{Turolla, R., Zampieri, L., Colpi, M., \& Treves, A. 1994, ApJ, 426, L35
(Paper I)}
\ref{Treves, A., \& Colpi, M. 1991, A\&A, 241, 107}
\ref{Verbunt, F., Belloni, T., Johnston, H.M., Van der Klis, M. \& Lewin,
W.H.G. 1993, A\&A, submitted}
\ref{Zel'dovich, Ya., \& Shakura, N. 1969, Soviet Astron.--AJ, 13, 175 (ZS)}

\vfill\eject
%
%----------------------------------------------------------------
%                        TABLE 1.
%----------------------------------------------------------------

\null
\vskip 1.5truecm
\centerline{Table 1}\medskip
\centerline{Characteristic Parameters for Selected Models}\bigskip\bigskip
$$\vbox{\tabskip=1em plus2em minus.5em
\halign to\hsize{#\hfil &\hfil # \hfil & \hfil # \hfil & \hfil # \hfil &
 \hfil # \hfil  & \hfil # \hfil & \hfil # \hfil &
\hfil # \hfil & \hfil # \hfil &\hfil # \hfil &\hfil # \hfil\cr
& & & $\displaystyle{{L_\infty}\over{L_{Edd}}}$ & $\Tg $ (keV) &
$\displaystyle
{{F_{>0.1}}\over{F^{bb}_{>0.1}}}^{\rm a}$ & $\gamma^{\rm b}$ &
$\gamma^{\rm c}$ & & &  \cr
\noalign{\smallskip}
%
% @@@@@@@@@@@@@@@@@@@@@@@@@@@@@@@@@@@@@@@@@@@@@@@@@@@@@@@@@@@@@@@@@@@@@
%
\noalign{\bigskip\medskip}
& & & $2.25\times 10^{-2}$ & 1.03 & 1.01 & 1.40 & -- & & &   \cr
& & & $10^{-3}$ & 0.53 & 1.03 & 1.56 & -- & & &  \cr
& & & $10^{-4}$ & 0.35 & 1.01 & 1.88 & -- & & &  \cr
& & & $10^{-5}$ & 0.24 & 1.02 & 2.21 & 2.06 & & &   \cr
& & & $10^{-6}$ & 0.15 & 1.07 & 2.44 & 2.31 & & &   \cr
& & & $10^{-7}$ & 0.09 & 1.34 & 2.64 & 2.55 & & &   \cr
 }}$$

\bigskip
\parindent 0.truept
$^{\rm a}$ Ratio of integrated flux to blackbody one above 0.1 keV,
for models with $y_0 = 20 \, {\rm g \, cm^{-2}}$.

\parindent 0.truept
$^{\rm b}$ Hardening ratio, defined in equation (12), for models with
$y_0 = 20 \, {\rm g \, cm^{-2}}$.

\parindent 0.truept
$^{\rm c}$ Hardening ratio for models with $y_0 = 5 \, {\rm g \, cm^{-2}}$.

%
%
% @@@@@@@@@@@@@@@@@@@@@@@@@@@@@@@@@@@@@@@@@@@@@@@@@@@@@@@@@@@@@@@@@@@@@
%
\vfill\eject

\beginsection FIGURE CAPTIONS

\refig{Figure 1.\quad Emergent spectra for $L_\infty = 2.25 \times
10^{-2}\, , 10^{-3}\, , 10^{-4}
\, L_{Edd}$ (full lines), together with the corresponding blackbody spectra
at the neutron star effective temperature (dashed lines).}
\medskip
\refig{Figure 2.\quad Same as in figure 1 for models with $L_\infty
= 10^{-5}\, , 10^{-6}\, , 10^{-7}
\, L_{Edd}$.}
\medskip
\refig{Figure 3.\quad Temperature vs. column density for different
values of $L_\infty/L_{Edd}$.}
\medskip
%\refig{Figure 4.\quad Same as in figure 1 for $y_0 = 5 \, {\rm g \,
%%cm^{-2}}$.}

\vfill\eject

\bye